\documentclass[11pt]{article}

\usepackage[utf8]{inputenc}
\usepackage[T1]{fontenc}
\usepackage{amsmath, amssymb}
\usepackage{geometry}
\usepackage{booktabs}
\usepackage{tabularx}
\usepackage{array}
\usepackage{enumitem}
\usepackage{hyperref}
\usepackage{tikz}
\usetikzlibrary{arrows.meta, positioning, shapes.geometric}

\setlist[itemize]{leftmargin=*}
\setlist[enumerate]{leftmargin=*}

\newcommand{\paperone}{A Predicate-Based Model for Computation over State Spaces}

\title{QDSV: A Semantic Problem Representation and Multi-Backend Execution Framework for Quantum-Oriented Computation}

\author{
Jaime Alexander Jim\'enez Lozano \\
QDSV Research Group \\
Colombia \\
\texttt{jaimeajl@qruba.site}
\and
Sebasti\'an Jim\'enez Giraldo \\
Universidad de los Andes \\
QDSV Research Group \\
Colombia \\
\texttt{s.jimenezg2345@uniandes.edu.co}
}

\date{}

\begin{document}

\maketitle

\begin{abstract}
Predicate-based computation over state spaces separates a problem specification from the backend that realizes it. A prior foundational paper introduced this abstraction formally as a pair $P=(S,C)$, where $S$ is a state space and $C:S\to\{0,1\}$ is a predicate whose satisfying states define the solution set. This work builds on the predicate-based model introduced in arXiv:2606.15027 and studies its realization as a semantic, multi-backend execution framework for quantum-oriented computation.

We describe how QDSV, QIntent, and Qruba connect a declarative problem intent to a structured semantic representation, realize that representation under heterogeneous backend constraints, and report execution trace outputs that separate model-level semantic outputs from backend-specific observations. The model can be realized through multiple execution modes, including backend realizations that do not require the original problem to be authored as a circuit and realizations that materialize circuit-compatible artifacts when required.

As a case study and backend validation, we evaluate EEG ictal/interictal classification using prepared signal features from the Bonn and Delhi datasets. The experiment compares classical machine-learning baselines, a circuit-first variational quantum classifier baseline, QDSV execution through simulator backends, and controlled IBM Quantum hardware runs. The results show that QDSV maintains consistency of semantic structure across multiple backend realizations without requiring circuit-first formulation as the primary abstraction.

The paper does not claim general quantum advantage or superiority over classical machine learning. Its contribution is a semantic execution validation showing how a problem-first representation can remain stable across simulator and hardware realizations while retaining execution trace outputs, without requiring the original problem to be stated as a circuit.
\end{abstract}

\section{Introduction}

Programming models often shape how users express problems before any execution backend is involved. Declarative and semantic approaches are valuable because they allow the structure of a problem to be stated before committing to a particular operational form. In quantum-oriented software, this separation becomes especially important: many workflows begin by asking the user to choose or design a circuit, feature map, ansatz, measurement scheme, and optimization loop. That circuit-first approach is powerful for expert users, but it can force a real problem into an operational representation before the structure of the problem has been made explicit. In structured data-centric and domain-specific computational settings, the relevant structure is often better expressed first through states, relations, constraints, prepared signals, and semantic criteria.

The foundational paper \emph{\paperone} introduced a predicate-based model in which a computational problem is specified as $P=(S,C)$: a state space $S$ and a predicate $C:S\to\{0,1\}$. It also introduced the notion of backend realization and semantic preservation. The purpose of the present paper is to validate whether this abstraction can remain stable across different backend constraints.

We present QDSV as a semantic execution model, QIntent as a restricted declarative surface language, and Qruba as a workflow environment used to construct and evaluate multi-backend executions. The paper focuses on the validation path from a declared problem to execution trace outputs:

\[
\text{intent} \rightarrow \text{intermediate semantic representation} \rightarrow \text{backend-constrained realization} \rightarrow \text{execution trace outputs}.
\]

QDSV is not primarily an algorithm-family selector. Its main execution path does not require the user to first choose Grover search, QAOA, VQE, a Hamiltonian formulation, a variational ansatz, or any other predefined quantum algorithmic template. Instead, QDSV starts from states, predicates, relations, prepared signals, semantic constraints, intended outputs, and evidence requirements. Algorithmic or circuit-level artifacts may be derived later when a target backend requires them, but they are not the primary representation of the problem.

This positioning is complementary to recent work on quantum languages, intermediate representations, and benchmark suites. OpenQASM~3 and QIR provide important representations for quantum programs and backend interoperability~\cite{cross2022,qir}. High-level modeling efforts such as Qmod aim to capture intent while delegating implementation details~\cite{vax2025}. Benchmark suites such as QASMBench and MQT Bench support systematic evaluation of circuits and tools across abstraction levels~\cite{li2022qasmbench,quetschlich2023mqtbench}. QDSV operates one layer earlier: it treats the problem representation itself as the primary object and derives backend-specific realizations only when needed.

The central claim is practical rather than asymptotic: a semantic problem representation can maintain consistency of the semantic structure of the problem across multiple backend realizations and can expose the execution trace outputs needed to interpret the result. In particular, when quantum hardware is noisy or probabilistic, a system should distinguish model-level semantic outputs from hardware-level observations and final platform outputs.

\paragraph{Contributions.}
This paper makes four contributions:
\begin{enumerate}
    \item It describes an abstract mapping from the predicate-based model $P=(S,C)$ to a structured QDSV problem representation.
    \item It connects that representation to QIntent, a restricted declarative surface for expressing problem intent without exposing arbitrary host-language execution.
    \item It presents a multi-backend validation framework for evaluating a stable semantic representation under different backend constraints.
    \item It evaluates the framework on an EEG ictal/interictal case study across simulator and IBM Quantum hardware realizations, including execution trace outputs and comparisons with classical and circuit-first VQC baselines.
\end{enumerate}

\paragraph{Scope.}
The paper is an initial semantic portability and backend validation paper. It does not claim that QDSV provides general quantum advantage, nor that it universally outperforms classical machine learning. The EEG experiment is used as an application case study for evaluating whether a declared semantic representation can be carried through different realization constraints while preserving interpretable execution trace outputs.

\section{From Predicate Model to Semantic Representation}

The foundational model defines a problem as $P=(S,C)$. In a backend validation setting, these mathematical objects are represented at a level that preserves their semantic role without requiring the paper to expose implementation-specific runtime mechanisms. Table~\ref{tab:mapping} summarizes the abstract correspondence used in this paper.

\begin{table}[h]
\centering
\small
\begin{tabularx}{\textwidth}{@{}p{0.18\textwidth}p{0.28\textwidth}X@{}}
\toprule
Formal object & Semantic representation & Role in validation \\
\midrule
$S$ & domain elements or finite state space & Defines the elements over which the problem is represented. \\
$C$ & semantic conditions over state attributes, relations, or signals & Defines which states satisfy the declared problem intent. \\
$\operatorname{Sol}(P)$ & semantic output set / solution set & Represents the states satisfying the semantic condition. \\
$R$ & backend-constrained realization & Maps the semantic representation under backend-specific constraints. \\
$\widehat{C}_R$ & realization-level evaluator & Evaluates or approximates the semantic condition under a concrete realization. \\
$E_R$ & execution trace outputs & Reports observable execution metadata, aggregate measures, reliability indicators, and warnings. \\
\bottomrule
\end{tabularx}
\caption{Abstract mapping from the predicate-based model to the QDSV semantic validation setting.}
\label{tab:mapping}
\end{table}

In tabular or data-centric problems, $S$ may be represented by indexed domain elements. A predicate $C$ may be represented by declared semantic conditions over prepared signals, relations, or state attributes. In combinatorial problems, $S$ may be represented by explicit finite domains and assignments. The implementation-specific details of how these representations are lowered into backend-specific artifacts are outside the scope of this paper.

\section{Semantic Representation and Multi-Backend Validation}

\begin{figure}[h]
\centering
\begin{tikzpicture}[
    node distance=0.9cm and 1.0cm,
    box/.style={rectangle, rounded corners, draw=black, align=center, minimum width=3.0cm, minimum height=0.85cm},
    modebox/.style={rectangle, rounded corners, draw=black, align=center, minimum width=4.0cm, minimum height=1.55cm},
    arrow/.style={-{Latex[length=2mm]}, thick}
]
\node[box] (intent) {Declared\\ intent};
\node[box, right=of intent] (ir) {Semantic\\ representation};
\node[box, right=of ir] (planner) {Backend-constrained\\ realization};
\node[modebox, below=of planner] (modes) {Realization modes\\[2pt]
\footnotesize Statevector-style realization\\
\footnotesize Circuit-compatible simulation\\
\footnotesize Quantum hardware realization};
\node[box, below=of modes] (evidence) {Execution trace\\ outputs};
\node[box, below=of evidence] (output) {Reported output\\ and evidence};

\draw[arrow] (intent) -- (ir);
\draw[arrow] (ir) -- (planner);
\draw[arrow] (planner) -- (modes);
\draw[arrow] (modes) -- (evidence);
\draw[arrow] (evidence) -- (output);
\end{tikzpicture}
\caption{High-level QDSV semantic validation flow.}
\label{fig:architecture}
\end{figure}
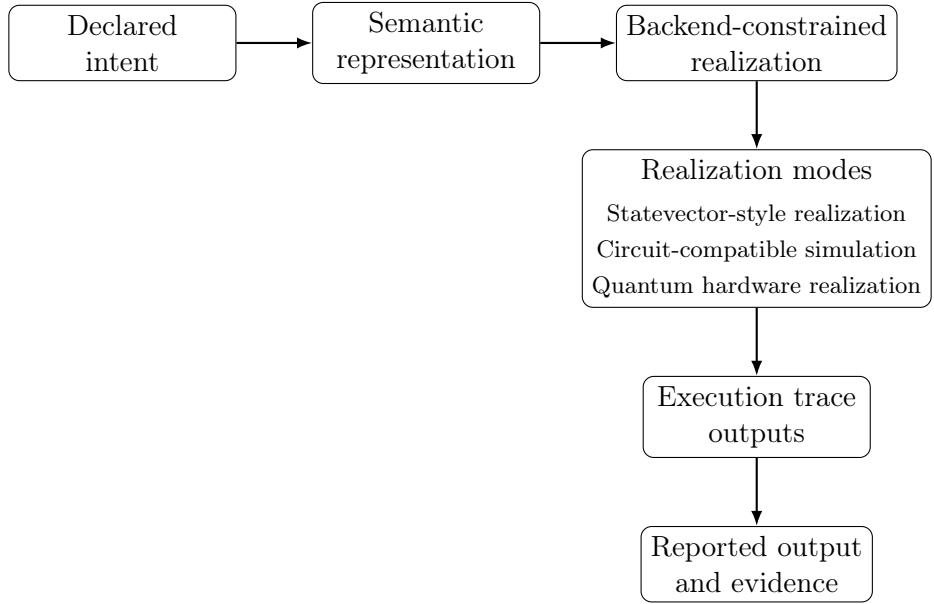

Figure~\ref{fig:architecture} summarizes the validation flow at a high level.

\subsection{QIntent}

QIntent is a restricted declarative surface for expressing problem intent. It is Python-like in syntax but is not general-purpose Python. Its purpose is to express domains, fields, predicates, relations, constraints, transformations, and domain-specific intents that can be compiled into a QDSV problem representation.

For example:

\begin{verbatim}
find_rows("candidate_index")
  .where("prepared_signal", ">=", threshold)
  .with_evidence(["backend", "solution_mass"])
\end{verbatim}

The intent is not executed as arbitrary Python. It is validated and compiled into a structured specification that can be inspected, executed, and audited.

\subsection{Intermediate Semantic Representation}

The intermediate semantic representation captures the contract between the declared problem and possible backend realizations. At the level needed for this paper, it includes:
\begin{itemize}
    \item state-space or domain definition;
    \item semantic condition family and representational structure;
    \item fields, prepared signals, relations, and criteria;
    \item backend constraints;
    \item requested output and execution trace fields;
    \item digests used for traceability.
\end{itemize}

The representation is intentionally narrower than a general programming language. Its purpose is to maintain consistency of the semantic structure needed for backend-constrained realization while avoiding arbitrary code execution. The implementation-specific lowering process is not described in this paper.

\section{Execution Modes}

QDSV supports several realization modes. Table~\ref{tab:modes} summarizes their intended roles at the abstraction level used in this paper.

\begin{table}[h]
\centering
\small
\begin{tabularx}{\textwidth}{@{}p{0.22\textwidth}p{0.28\textwidth}X@{}}
\toprule
Mode & Role & Circuit requirement \\
\midrule
Semantic evaluation & Evaluates declared semantic conditions over the prepared semantic representation. & No user-written circuit. \\
Statevector-style realization & Represents the solution structure through a statevector-oriented backend constraint. & Circuit materialization is not the primary abstraction. \\
Circuit-compatible simulation & Uses a circuit-compatible realization when a simulator requires that form. & Circuit artifact is derived from the semantic representation. \\
Hardware realization & Uses a backend-compatible realization on quantum hardware for controlled subsets. & Circuit-compatible artifact may be required by the backend. \\
\bottomrule
\end{tabularx}
\caption{QDSV realization modes and their relation to circuit materialization.}
\label{tab:modes}
\end{table}

This distinction is central. QDSV does not treat the circuit as the primary problem representation. When a backend requires a circuit-compatible form, that form is treated as a realization artifact derived from the semantic representation. When a backend can support a statevector-style realization, the original problem need not be forced into a circuit template as the starting point.

Consequently, QDSV should not be read as a wrapper around standard quantum algorithms. It may interoperate with known algorithmic forms when useful, but the model-level object remains the semantic problem representation: states, predicates, relations, signals, solution structures, and evidence.

\section{Execution Trace Outputs and Traceability}

For practical quantum-oriented execution, a result should not be reported only as a selected state, scalar value, or final label. It should also include execution trace outputs describing how the result was realized. QDSV reports this information through an evidence record associated with each realization.

This emphasis on evidence and traceability is consistent with broader concerns in quantum software engineering, where testing, analysis, unexpected behavior, and reliability remain active challenges across quantum programming stacks~\cite{paltenghi2024survey}.

The evidence record includes:
\begin{itemize}
    \item backend identity and execution path;
    \item solution count and solution mass;
    \item shots, counts, probabilities, and ranks when available;
    \item model-level semantic output indicators;
    \item backend reconstruction indicators when available;
    \item reliability status and acceptance indicators;
    \item digests and warnings.
\end{itemize}

QDSV reports three complementary layers that connect the semantic representation with backend execution evidence:

\begin{center}
\small
\begin{tabularx}{\textwidth}{@{}p{0.30\textwidth}X@{}}
\toprule
Layer & Meaning \\
\midrule
Semantic output & Output implied by the semantic representation and model-level condition. \\
Backend evidence & Candidate-level or aggregate observations reconstructed from backend execution when available. \\
Reported output & Platform-level output that preserves the semantic result while reporting backend reliability. \\
\bottomrule
\end{tabularx}
\end{center}

This separation is important on noisy hardware. A hardware run may confirm aggregate solution mass while not providing enough candidate-level evidence for every state. In such cases, the system should not silently replace model-level semantic outputs with unreliable backend reconstructions.

\subsection{Reliability for Solution Sets}

Early hardware policies can incorrectly focus on the probability of a single dominant state. However, many QDSV formulations produce solution structures rather than a single winner. When the expected output is a set of marked states, reliability should be evaluated using aggregate solution mass rather than only dominant-state probability.

Let $M\subseteq S$ be the set of states selected by the semantic predicate. A solution-set reliability check compares the observed hardware mass
\[
\widehat{p}(M)=\frac{\text{shots observed in }M}{\text{total shots}}
\]
against the expected or simulated mass under the selected realization. This is more appropriate than requiring a single state to dominate when the correct solution contains many states.

\section{Case Study and Backend Validation: EEG Ictal/Interictal Classification}

We evaluate QDSV on EEG ictal/interictal classification using prepared signal features from the Bonn and Delhi datasets. The experiment is designed as both an application case study and a backend validation. The goal is not to claim medical deployment readiness, but to test whether a semantic problem representation can maintain consistency of the semantic structure of the problem across different backend realizations.

\subsection{Datasets}

The Bonn experiment uses a binary setting with interictal and ictal samples. The prepared dataset contains 300 rows: 200 interictal and 100 ictal. The Delhi experiment uses 100 rows in the binary setting: 50 interictal and 50 ictal. Preictal samples were excluded from the binary evaluation.

Each row contains extracted numerical EEG features. The feature extraction includes Hjorth-style features, entropy, statistical features, zero crossings, and wavelet-derived energy and moment features. The QDSV input uses prepared, normalized, and oriented signal scores.

\subsection{QDSV Formulation}

In the EEG validation, the operational Qruba component is the \emph{ScoreModel} node. In this experimental setting, the node takes prepared values or signals, maps them into semantic factors, applies a declared condition, and produces state-level outputs with associated execution trace outputs. The node is not merely a statistical post-processing step: it is the system component that turns prepared evidence into a QDSV semantic condition over candidate states and then exports model-level semantic outputs, backend observations, and reported output columns.

At a high level, the workflow is:
\[
\text{prepared signals} \rightarrow \text{semantic factors} \rightarrow
\text{predicate/evaluation} \rightarrow \text{solution structure} \rightarrow \text{evidence}.
\]
This paper reports the observable contract and experimental behavior of that workflow, not proprietary internal implementation details.

For BONN, the recommended QDSV semantic factors used in this case study were:
\begin{itemize}
    \item \texttt{dwt\_cD3\_std\_score}
    \item \texttt{std\_score}
    \item \texttt{activity\_score}
    \item \texttt{energy\_score}
    \item \texttt{dwt\_cD2\_std\_score}
    \item \texttt{complexity\_score}
\end{itemize}

For DELHI, the recommended QDSV semantic factors used in this case study were:
\begin{itemize}
    \item \texttt{activity\_score}
    \item \texttt{energy\_score}
    \item \texttt{std\_score}
    \item \texttt{dwt\_cA3\_energy\_score}
    \item \texttt{dwt\_cD3\_energy\_score}
    \item \texttt{dwt\_cD2\_energy\_score}
\end{itemize}

Labels and clinical classes were used only for validation, not as model inputs.

\subsection{Baselines}

We compare against two families of baselines:
\begin{itemize}
    \item classical machine-learning baselines using Decision Tree and Random Forest over feature reductions such as ALL, PCA, SVD, NMF, and UMAP;
    \item circuit-first VQC baselines using angle encodings and RealAmplitudes-style circuits.
\end{itemize}

The classical baselines represent a strong non-quantum reference. The VQC baselines represent the circuit-first route in which the data must be reduced and encoded into a fixed circuit form before measurement.

\section{Results}

The reported results use different protocols because they answer different questions. Table~\ref{tab:protocols} summarizes the role of each result family. The classical and VQC baselines are predictive baselines. The QDSV simulator and hardware runs are backend-validation experiments intended to evaluate semantic portability, execution trace outputs, and reliability behavior under simulator and hardware constraints.

\begin{table}[h]
\centering
\small
\begin{tabularx}{\textwidth}{@{}p{0.22\textwidth}p{0.28\textwidth}X@{}}
\toprule
Result family & Protocol & Primary role \\
\midrule
Classical baselines & Repeated holdout & Predictive reference using classical machine-learning models. \\
Circuit-first VQC baselines & Repeated and focused VQC runs & Quantum circuit-first reference under feature reduction and circuit encoding. \\
QDSV simulator validation & Full-dataset operational validation & Backend consistency check for the same semantic formulation across simulator realizations. \\
IBM hardware validation & Controlled hardware subsets & Hardware evidence, solution-set reliability, and semantic/backend output separation. \\
\bottomrule
\end{tabularx}
\caption{Experimental result families and their protocols. The protocols are not identical and should not be interpreted as a single leaderboard.}
\label{tab:protocols}
\end{table}

\subsection{Classical and Circuit-First Baselines}

Table~\ref{tab:baselines} summarizes the strongest observed classical and circuit-first VQC baselines from the repeated holdout and focused VQC runs.

\begin{table}[h]
\centering
\small
\begin{tabularx}{\textwidth}{@{}p{0.16\textwidth}p{0.25\textwidth}p{0.14\textwidth}p{0.14\textwidth}p{0.14\textwidth}X@{}}
\toprule
Dataset & Method & Accuracy & Precision & Recall & F1 ictal \\
\midrule
BONN & RandomForest + NMF(10) & 0.9815 & 0.9780 & 0.9667 & 0.9718 \\
DELHI & DecisionTree/RandomForest + ALL(29) & 1.0000 & 1.0000 & 1.0000 & 1.0000 \\
BONN & VQC focused, NMF(7), angle\_ry & 0.6267 & 0.5404 & 0.8533 & 0.6344 \\
DELHI & VQC focused, PCA/SVD(7), angle\_ry & 0.6200 & 0.6074 & 0.8400 & 0.6878 \\
\bottomrule
\end{tabularx}
\caption{Representative classical and circuit-first VQC baselines. Values are means from repeated runs where applicable.}
\label{tab:baselines}
\end{table}

The VQC baseline obtains relatively high ictal recall but substantially lower precision and F1 than the classical baseline. This suggests that, in the evaluated configuration, the reduction, encoding, ansatz, and measurement route introduced additional representational translation steps before backend execution.

\subsection{QDSV Simulator Validation}

Table~\ref{tab:sim} shows full-dataset QDSV operational validation on BONN using the QuEST-based semantic/statevector path and Aer. This is a backend consistency check over the prepared semantic formulation, not a train/test predictive evaluation.

\begin{table}[h]
\centering
\small
\begin{tabularx}{\textwidth}{@{}p{0.15\textwidth}p{0.12\textwidth}p{0.12\textwidth}p{0.13\textwidth}p{0.13\textwidth}p{0.13\textwidth}X@{}}
\toprule
Dataset & Backend & Rows & Accuracy & Precision & Recall & F1 ictal \\
\midrule
BONN & QuEST & 300 & 0.9800 & 0.9700 & 0.9700 & 0.9700 \\
BONN & Aer & 300 & 0.9800 & 0.9700 & 0.9700 & 0.9700 \\
\bottomrule
\end{tabularx}
\caption{Full-dataset QDSV simulator validation on BONN. This is an operational backend validation, not a repeated holdout predictive protocol.}
\label{tab:sim}
\end{table}

The same semantic formulation produced matching classification metrics on the QuEST-based semantic/statevector path and Aer. This supports the validation goal of checking whether the semantic representation behaves consistently across simulator backends while allowing different realization requirements. In particular, the QuEST path is used here as a statevector-style realization, whereas Aer represents a circuit-compatible realization path.

\subsection{IBM Quantum Hardware Validation}

IBM Quantum hardware runs were performed on controlled subsets due to hardware cost, queueing, and shot limitations. The hardware experiments are not treated as full-dataset clinical validation. They are used to validate execution evidence, solution-set reliability, and separation between model-level semantic outputs and backend observations.

Table~\ref{tab:ibm} summarizes the controlled IBM Marrakesh hardware runs selected for reporting. The DELHI runs correspond to the latest solution-set reliability policy and were accepted by that policy. The BONN runs are included as additional hardware evidence under report-only reliability.

\begin{table}[h]
\centering
\small
\begin{tabularx}{\textwidth}{@{}p{0.11\textwidth}p{0.08\textwidth}p{0.09\textwidth}p{0.11\textwidth}p{0.10\textwidth}p{0.10\textwidth}p{0.09\textwidth}p{0.09\textwidth}X@{}}
\toprule
Dataset & Rows & Selected & Sol. mass & Accuracy & Precision & Recall & F1 ictal & Reliability \\
\midrule
DELHI & 64 & 33 & 0.5068 & 0.9844 & 0.9697 & 1.0000 & 0.9846 & accepted \\
DELHI & 32 & 18 & 0.5664 & 0.9375 & 0.8889 & 1.0000 & 0.9412 & accepted \\
BONN & 64 & 32 & 0.4863 & 1.0000 & 1.0000 & 1.0000 & 1.0000 & reported \\
BONN & 64 & 31 & 0.4971 & 0.9844 & 1.0000 & 0.9688 & 0.9841 & reported \\
\bottomrule
\end{tabularx}
\caption{Controlled IBM Marrakesh hardware runs. DELHI runs use accepted solution-set reliability; BONN runs are report-only additional hardware evidence.}
\label{tab:ibm}
\end{table}

The DELHI 64-row IBM run achieved F1 ictal $0.9846$ with recall $1.0$, detecting all ictal rows in the subset with one false positive. The DELHI 32-row IBM run also achieved recall $1.0$, with F1 ictal $0.9412$. These subset-level results are not intended as a head-to-head claim of superiority over the full classical baselines. Rather, they show that the QDSV semantic formulation retains strong ictal discrimination while producing IBM hardware execution trace outputs, and that it does not require circuit-first formulation as the primary abstraction in this case study.

\subsection{Solution Mass and Reliability}

The DELHI 64-row IBM run selected 33 rows with observed solution mass approximately $0.5068$. The DELHI 32-row IBM run selected 18 rows with observed solution mass approximately $0.5664$. Both runs were classified as reliable under the solution-set reliability policy.

This is important because the correct output is not a single dominant state. The expected structure is a set of selected candidates. Therefore, aggregate solution mass is a more appropriate reliability signal than dominant-state probability alone.

\subsection{Secondary Semantic Reconfiguration Case Study: Loan Approval Data}

As an additional domain check, QDSV was exercised on a loan-approval dataset to evaluate whether the same data representation could support different semantic intents under the same validation flow.

The dataset was evaluated under two intents. The first intent represented an expert-defined low-risk formulation using conservative financial criteria. The second intent represented an approval-likelihood formulation aligned with historical approval labels. In the approval-likelihood run, QDSV returned 31 out of 999 candidates, and all returned candidates were historically approved. This corresponds to a highly selective, high-confidence subset rather than an exhaustive credit-risk model.

The QuEST execution concentrated the probability distribution on the selected subset (solution mass $\approx 0.99999$), indicating that the backend execution was strongly aligned with the semantic solution set defined by the QDSV formulation used in this case. This value is interpreted as backend evidence for the selected solution set, not as a standalone predictive metric.

The significance of this experiment is that the semantic intent could be reconfigured over the same dataset while retaining the QDSV representation, execution trace outputs, and backend realization path. This illustrates how different semantic formulations can be expressed and executed over a stable problem representation.

\section{Discussion}

The reported experiments provide preliminary evidence supporting three observations.

\paragraph{First, QDSV maintains consistency of semantic structure across backend realizations.}
The BONN full-dataset QuEST and Aer runs produced consistent metrics. The same semantic representation could be used across multiple backend realizations rather than redesigning the problem for each backend. This is especially important because the QuEST path does not require the user-facing problem to begin as a circuit, while Aer and IBM require circuit-compatible materializations.

\paragraph{Second, QDSV does not require circuit-first formulation as the primary abstraction.}
The VQC baseline achieved F1 ictal around $0.63$ on BONN and $0.69$ on DELHI. QDSV reached F1 ictal $0.97$ on BONN simulator validation and $0.9846$ on the DELHI 64-row IBM hardware subset. This does not imply quantum advantage, but it suggests that reducing representational translation steps before backend realization can be beneficial in this case study.

\paragraph{Third, hardware execution trace outputs must be interpreted carefully.}
On IBM hardware, a high-level accuracy score can be misleading if the system does not separate model-level semantic outputs from backend observations. QDSV reports semantic outputs, backend reconstruction where available, reported outputs, and reliability status. This makes it possible to distinguish model-level structure from physical execution trace outputs.

\section{Limitations}

The evaluation has several limitations.
\begin{itemize}
    \item The EEG case study is not a clinical validation study.
    \item The reported result families use different protocols: repeated holdout for classical and VQC baselines, full-dataset operational validation for QDSV simulator runs, and controlled subsets for IBM hardware runs.
    \item IBM hardware runs were performed on controlled subsets, not full datasets.
    \item Hardware results depend on backend availability, calibration state, queueing, shots, and noise.
    \item The primary case study uses binary ictal/interictal classification; broader validation should include additional domains and problem forms.
    \item Circuit-compatible realization may introduce overhead not captured by high-level semantic metrics.
    \item The current paper reports execution trace outputs but does not provide a full formal semantics for QIntent or a complete implementation specification of QDSV.
\end{itemize}

\section{Conclusion}

This paper presented an initial multi-backend validation of the predicate-based model introduced in \emph{\paperone}. QDSV maps state spaces and predicates into semantic problem representations, realizes them across multiple backend constraints, and reports execution trace outputs.

The EEG case study shows that QDSV can maintain consistency of semantic structure across a QuEST-based semantic/statevector route, Aer, and IBM Quantum hardware runs. The evaluated workflow does not require circuit-first formulation as the primary abstraction and remains close to strong classical ML references on the evaluated settings, while retaining explicit execution trace outputs and reliability reporting.

Future work includes further formalizing QDSV semantic representations, expanding QIntent semantics, adding more benchmark domains, reporting larger hardware experiments, and studying semantic preservation under approximate and noisy backend realizations.

\section*{Data and Reproducibility}

The experiments use locally prepared EEG feature tables and QDSV/Qruba outputs. A public reproducibility package should include feature extraction scripts, prepared QDSV inputs, exported QDSV outputs, evaluation notebooks, and backend metadata. Sensitive credentials and private runtime internals are not required for reproducing the reported metrics.

\section*{Author Contributions}

Jaime Alexander Jim\'enez Lozano conceived the QDSV semantic validation framework, developed the QDSV/QIntent/Qruba prototype, designed the backend validation experiments, and led the manuscript writing.

Sebasti\'an Jim\'enez Giraldo contributed to conceptual review, experimental interpretation, validation of the case-study framing, and manuscript revision.

\end{document}